# Plasmonic Nano-arrays for Enhanced Photoemission and Photodetection

**SHIVA PILTAN,**[1,*] **DAN SIEVENPIPER**[1]

[1] *Electrical and Computer Engineering Department, University of California San Diego, La Jolla, California 92093, USA*
*\*Corresponding author: spiltan@eng.ucsd.edu*

**Efficient conversion of photons to electrical energy has a wide variety of applications including imaging, energy harvesting, and infrared detection. The coupling of electromagnetic radiation to free electron oscillations at a metal interface results in enhanced electric fields tightly confined to the surface. Taking advantage of this nonlinear light-matter interaction, this work presents resonant surfaces optimized for combining electrical and photonic excitations in order to liberate electrons in a vacuum-channel device for applications ranging from enhanced photoemission to infrared photodetection.**

***OCIS codes:*** *(250.5403) Plasmonics; (250.0040) Detectors; (040.3060) Infrared; (040.5160) Photodetectors; (240.6680) Surface plasmons.*

## 1. INTRODUCTION

In order to liberate cold, bound electrons from metal surfaces it is necessary to provide sufficient energy to the electrons to promote to high energy level states and overcome the potential barrier, or to modify the barrier using DC or AC external fields in mechanisms such as field emission, photo-assisted and optical field emission, and photoemission. The input power levels required to create the desired combination of carrier state and barrier shape can be significantly moderated using the nonlinear light-matter interaction between metals and incoming optical radiation in plasmonic frequencies. The coupling of electromagnetic radiation to free electron oscillations at a metal interface and its consequent properties including enhanced optical near-field have drawn significant interest to the field of plasmonics and its applications. Some of the most widespread demonstrated applications of this nano-scale light-matter interaction include surface-enhanced Raman spectroscopy (SERS) [1,2], plasmonic color pixels for CMOS compatible imaging and bio-sensing [3-5], hot electron photo-electrochemical and photovoltaic devices and photodetectors [6-9], optical antennas [10,11], plasmonic integrated circuits [12,13], and metamaterials [14,15]. We have taken advantage of the plasmonic field enhancement in resonant metallic arrays of unit-cells in order to address the carrier generation challenge in a variety of vacuum microelectronic devices for applications including enhanced photoemission and photodetection. The geometrical dependence of the resonance frequency and its consequent localized field enhancement offers an important degree of freedom for optimizing the device for a spectrum of frequencies.

The addition of optical excitation to static external fields can interact with the system by both modulating the barrier and increasing the initial energy of electrons due to photon absorption [16-19]. The dominant mechanism for relatively weak optical fields is either photo-assisted field emission or photoemission. In the photo-assisted field emission process, electron energy is enhanced to a non-equilibrium distribution by absorption of one or more photons of frequency ω and tunnels through thereafter. In the photoemission process, the electron absorbs a sufficient number of photons to travel above the barrier. If the laser intensity is sufficiently large, the potential barrier becomes narrow enough during part of the optical cycle for the electrons to tunnel through directly from the Fermi level and the dominant emission mechanism is called optical field emission.

Fig. 1(a) shows the schematic of the periodic surface supporting local surface plasmon resonances in the near-infrared range. There are two terminals serving as the feed for the DC excitation and every other row of the array is connected to one terminal. Therefore, a vacuum channel with 50 nm width is formed between the tips of each triangular unit-cell on two adjacent rows. Under normal conditions the vacuum gap prevents significant conductance between the two sharp terminals until the DC bias voltage reaches levels up to 100s of volts. Due to the sharp tip geometry of unit-cells the sensitivity of the surface to optical excitation is enhanced and electron emission and acceleration occurs at lower power intensities. Significant photocurrents are observed using combination of on the order of $1\,\frac{\text{W}}{\text{cm}^2}$ optical power density and 10 volts DC excitation on the surface.

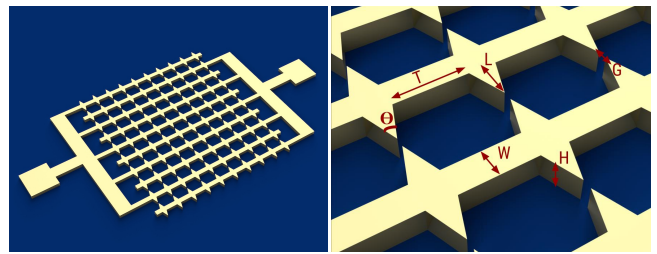

**Fig. 1.** Plasmon-induced photoemission vacuum-channel device. (a) Resonant array emits electrons under a CW laser illumination and a DC bias. (b) Multiple unit-cells of the triangular array and the simulated geometric parameters, T= 400 nm, L= 200 nm, W= 150 nm, θ= 40 degrees, and H= 130 nm.

The geometry of the proposed structure is optimized to reach maximized electric field enhancement at the tip of the sharp triangular unit-cell for a normally incident wave polarized in direction of the tip axis and with a wavelength of 785 nm. Each geometrical dimension is parametrically swept to maximize the electric field at the tip and the process is iterated to ensure optimized dimensions for a specific incident electric field wavelength. The structure is simulated using the Ansys HFSS electromagnetic solver based on the finite-element method and the Johnson-Christy model for the complex dielectric constant properties of gold [20]. The optimized unit-cell dimensions shown in Fig. 1(b) are T= 400 nm, L= 200 nm, W= 150 nm, θ= 40 degrees, and H= 130 nm.

Field enhancement at any point is defined as the ratio of the electric field at that point to incident field. Fig. 2(a) demonstrates the field enhancement value as a function of frequency for two points, on the substrate and 65 nm above silicon dioxide substrate, both on the tip edge. Fig. 2(b) shows the electric field profile at λ= 785 nm on the side interface of the half unit-cell. The maximized enhancement regions correspond to the top and bottom corners on the edge of tip. This is mainly due to existence of sharper features in top and bottom corners associated with higher electric field enhancements and consequently more significant electron emission. The simulated field profiles show enhancement factors as large as 700 on the corners at λ= 785 nm (382 THz).

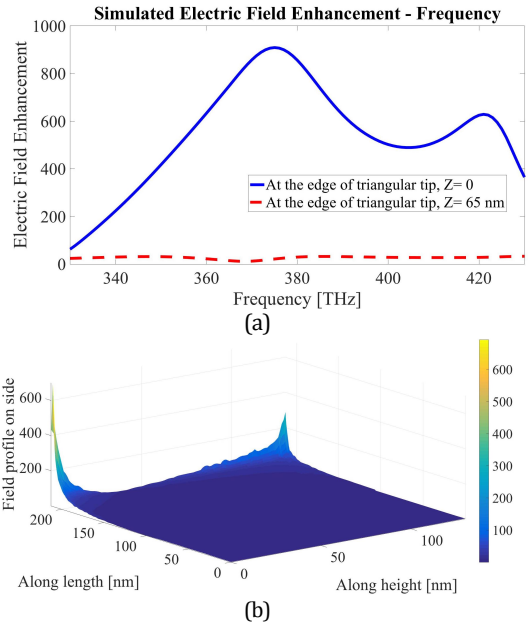

**Fig. 2.** (a) Simulated field enhancement as a function of frequency for final optimized dimensions, solid (blue) at the edge of the triangular tip on the substrate, dashed (red) at the edge of the triangular tip 65 nm above substrate. (b) Simulated field enhancement profile on the side interface of a triangular half unit-cell upon incidence of plane wave at 785 nm polarized along the tip axis.

A 40x40 layout of the periodic unit-cells was fabricated on silicon wafers with a 280 nm thermally grown $SiO_2$ layer to minimize leakage currents through substrate. The silicon wafers had a resistivity above 10000 Ω.cm and the thickness of $SiO_2$ layer is sufficient for proper isolation [21]. We do not expect thermal expansion in our experiment since the electrodes are mechanically coupled to the substrate. Thermal voltages are not significant because of the symmetric design and the minor temperature gradients between the electrodes [22]. The thermally enhanced field emission process is not of critical contribution for static fields below on the order of 1 GV/m as discussed by Kealhofer et al. for hafnium carbide tips [23]. The layout was exposed using e-beam lithography and subsequently metallized using e-beam evaporation of a 5 nm chromium adhesion layer and a 125 nm gold layer. Fig. 3 shows scanning electron microscope images of the fabricated arrays.

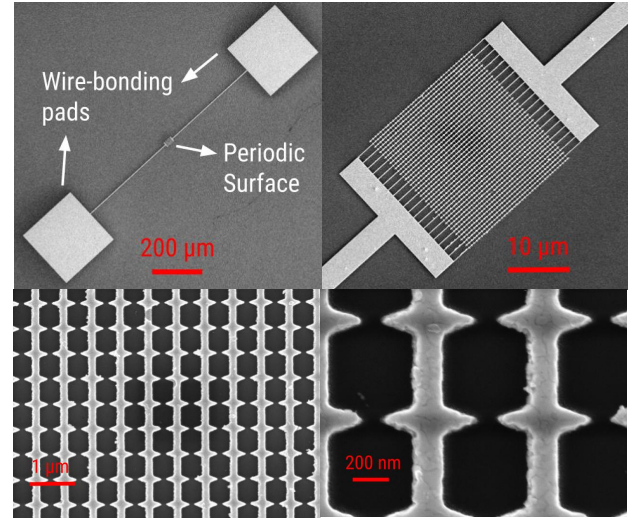

**Fig. 3.** Scanning electron microscope images of periodic surface after fabrication.

The measurement setup consists of a tunable Ti:Sapphire continuous-wave laser with wavelength centered at 785 nm and beam radius of 0.475 mm pumped with a 10 watt green semiconductor laser. The laser beam is first sampled for power and wavelength measurements to be done using a silicon photo-detector and spectrometer. The beam is then focused on the sample inside a vacuum chamber pumped down to 0.1 mTorr. The measured current-voltage curves for various radiation power densities are shown in Fig. 4(a). Fig. 4(b) demonstrates the current as a function of optical power density for fixed DC voltages. In all measurements for each incident optical power density, the DC voltage applied between terminals is swept from -12 V to 12 V and the current is captured using a source-meter. Figures corresponding to current as a function of laser power intensity are then extracted using the data for a specific DC voltage from the same measurements. For low laser power densities, the DC electric field is not sufficient for the electrons to overcome the potential barrier as expected; however, it is shown that the emitted current increases significantly as the optical illumination intensifies to levels as low as 10s of $\frac{W}{cm^2}$.

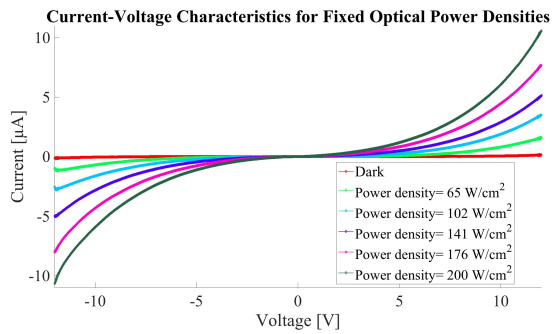

(a)

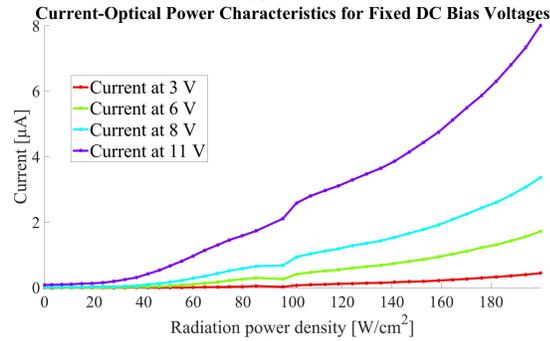

(b)

**Fig. 4.** (a) Measured current-voltage characteristics for the triangular sharp tip array for fixed radiation intensities centered at 785 nm (0, 65, 102, 141, 176, and 200 $\frac{W}{cm^2}$ shown). (b) Measured current-optical power characteristics for the triangular sharp tip array for fixed DC bias voltages (3, 6, 8, and 11 V shown).

Since every single cell contributes to the photocurrent, it is expected for the current to scale with the area of the array. Scaling is an important property because it can compensate for low effective quantum efficiency and it introduces new applications such as alternative solutions for energy harvesting. A scaled array will have a larger current handling capacity and it will increase the damage threshold of the device. In order to confirm this behavior, we fabricated an 80x80 periodic array of the same structure and compared the measured photocurrent for identical static fields and optical intensities in the scaled array with the original 40x40 size. The resulting photocurrent as a function of optical power density for multiple voltages is shown in Fig. 5. The scaling behavior of photocurrent with surface area for a variety of DC and optical powers is clearly shown. Since the array size in each dimension is doubled, the emitted current in the 80x80 element array is approximately 4 times bigger than the 40x40 surface as expected. In other words the emitted current normalized to the surface area of the array in both 40x40 and 80x80 scales is very similar. As an example, current per unit surface area at a combination of 6.5 V DC voltage and 70 $\frac{W}{cm^2}$ optical intensity is 430 $\frac{A}{m^2}$ in the 40x40 array compared to 465 $\frac{A}{m^2}$ in the 80x80 array.

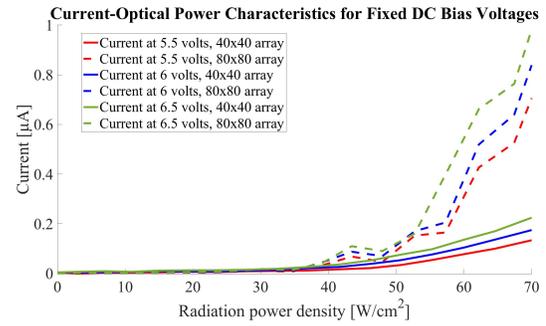

**Fig. 5.** Measured current-optical power characteristics for fixed DC bias voltages (5.5, 6, and 6.5 V shown) for 40x40 (solid) and 80x80 (dashed) element surface. The larger array emits approximately 4 times higher photocurrent for similar electrical and optical input powers.

There are three most widely used approaches to theoretically understand the phenomena concerned with nonlinear interaction of laser and matter, the strong-field approximation [24], the semi-classical approach [25], and the direct solution to the time-dependent Schrödinger equation discussed by Zhang and Lau for a modulated triangular potential barrier [26]. Plotting the current-voltage characteristic curves in Fowler-Nordheim [27] format gives useful insight to the electron emission processes. Fig. 6 demonstrates the measured current-voltage characteristics in this format. The linear behavior confirms emission of electrons through the free space rather than leakage currents through substrate. Other analytical methods for characterizing the surface properties of electric field include Tien-Gordon approach [28, 29] utilized by Ward et al. [22] for characterizing optical rectification in plasmonic nano-gaps, which is outside the scope of this paper.

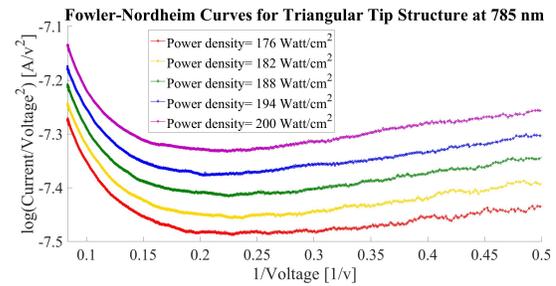

**Fig. 6.** Measured $\log\left(\frac{I}{V^2}\right) - \frac{1}{V}$ characteristic curves for fixed radiation intensities centered at 785 nm (176, 182, 188, 194, and 200 $\frac{W}{cm^2}$ shown).

The geometric degree of freedom for tailoring the resonance frequency and field enhancement allows for optimizing the device for applications in the infrared spectrum as well. Infrared radiation plays an important role in providing access to physical phenomena otherwise invisible to naked eye used in thermal imaging, environmental monitoring, medical imaging, and remote sensing [30-32]. The thermal infrared radiation emitted by many terrestrial objects is mainly centered in two atmospheric wavelength windows, the middle wavelength infrared (MWIR) region in 3-5 µm band, and the long wavelength infrared (LWIR) region in 8–14 µm band, since maximum emissivity of gray objects in room temperature is centered at wavelengths close to 10 µm [33].

Infrared detectors can be classified in two main categories based on the detection mechanism [33]. The absorbed radiation can be detected in material either by interacting with electrons or by changing the temperature of the sensor. Photon detectors are based on the former mechanism and thermal detectors operate due to the latter. The advantages of photon detectors compared to thermal detectors include higher signal to noise ratio and short response time; however, this often requires cryogenic cooling which adds cost and weight and it is not convenient for all applications. Thermal detectors usually consist of a temperature-dependent sensing element that is thermally isolated from surrounding structures and it is usually built in a suspended configuration. Thermal detectors are relatively cheap and convenient to use because they typically operate at room temperature; however, their performance depends highly on the detector element's heat transfer characteristics. Photon detectors provide superior performance as long as the temperature of the device can be maintained low. Otherwise, thermally generated charge carriers undermine the signal to noise ratio. High-temperature operation of infrared detectors is therefore one of the main challenges limiting their applicability.

Among all types of photodetectors, the variable band gap HgCdTe sensors remain one of the most commonly used [34]. HgCdTe provides high spectral tunability and it is widely used in applications spanning in short, mid, long, and very long wavelength infrared bands. The most significant challenge associated with HgCdTe sensors relates to the difficulties in processing of this material and achieving strong Hg-Te bonds [34]. Maintaining performance of these detectors working under photoconductive, photovoltaic, and metal-insulator-semiconductor designs in high operational temperatures is another drawback to be addressed.

Nanostructures offer unique access to geometrical and optoelectronic properties not available otherwise [35]. Nano-devices such as quantum wells, carbon nanotubes, quantum dots, and graphene-based nanomaterials have been used for enhancing performance metrics of infrared photo-detectors. Plasmon-enhanced photodetectors [36] can compensate for the optical losses in metals by addressing overall challenges in flexibility, tunability, operation speed, and power consumption. Here, we used a similar approach to take advantage of the size, geometry, and material of the designed nano-array in order to target infrared photodetection applications.

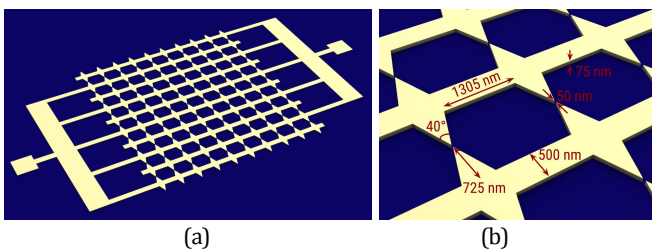

**Fig. 7.** (a) Resonant surface for maximized field enhancement at 25 THz. (b) Simulated optimized dimensions.

The idea behind the design is very similar to that discussed previously for the sharp tip periodic array of electron emitters at λ= 785 nm. Fig. 7(a) shows the modified geometry combining electrical and optical excitations at 12 μm. In order to keep the fast response and maximized enhancement factor at the surface, we kept the vacuum gap width between the unit-cells on adjacent rows at 50 nm. The rest of the geometrical parameters were optimized for maximizing the electric field at the tip of the triangular cell upon excitation with a plane wave at 25 THz (12 μm) incident normally and polarized in the direction of the tip axis. The final dimensions based on the simulations are shown in Fig. 7(b). Fig. 8(a) demonstrates the simulated field enhancement factor as a function of frequency at the edge of the tip on the substrate and 37.5 nm above substrate. Fabrication was done using e-beam lithography and e-beam evaporation. Fig. 8(b) shows scanning electron microscope image of samples after fabrication.

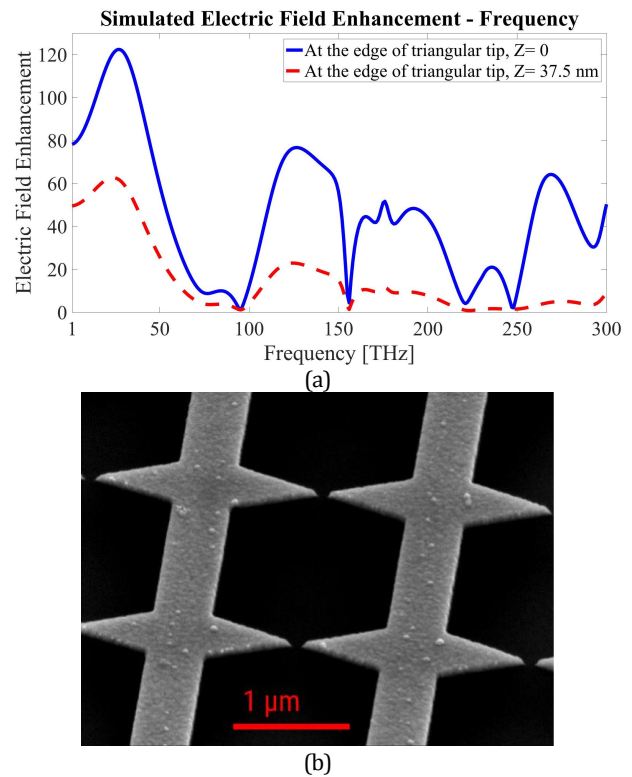

**Fig. 8.** (a) Simulated electric field enhancement for optimized array for 25 THz at the edge of the tip. Solid (blue) shows field enhancement on the substrate and dashed (red) is 37.5 nm above substrate. (b) Scanning electron microscope image of fabricated samples.

The current-voltage characteristics of the device were measured using emission from a silicon nitride furnace igniter at 1300 °C as infrared excitation. Radiation was focused on the sample using a ZnSe meniscus lens with focal length of 63.5 mm and diameter of 28 mm. Fig. 9(a) demonstrates the measured current-voltage curves and resulting specific detectivity is shown in Fig. 9(b). Specific detectivity is the main figure of merit for characterizing normalized signal to noise performance of infrared detectors and it is defined as:

$$D^* = \frac{\sqrt{A\Delta f}}{NEP} \quad (1)$$

where A is the detector area, Δf is the measurement bandwidth, and NEP is the noise equivalent power. Noise equivalent power is defined as the optical power needed to induce a current equivalent to the noise current of detector and it can be written as the ratio of the noise current to responsivity. Noise current $I_n$, in our device is the sum of shot noise from the photocurrent $I_{ph}$, and the Johnson noise from detector resistance R [37,38] and it can be expressed as:

$$I_n = \sqrt{(2qI_{ph} + \frac{4K_BT}{R})\Delta f} \quad (2)$$

Calculations based on unity emissivity of the thermal emitter result in an upper bound on emitted radiance and lower bound on detectivity; therefore, we assumed emissivity as unity.

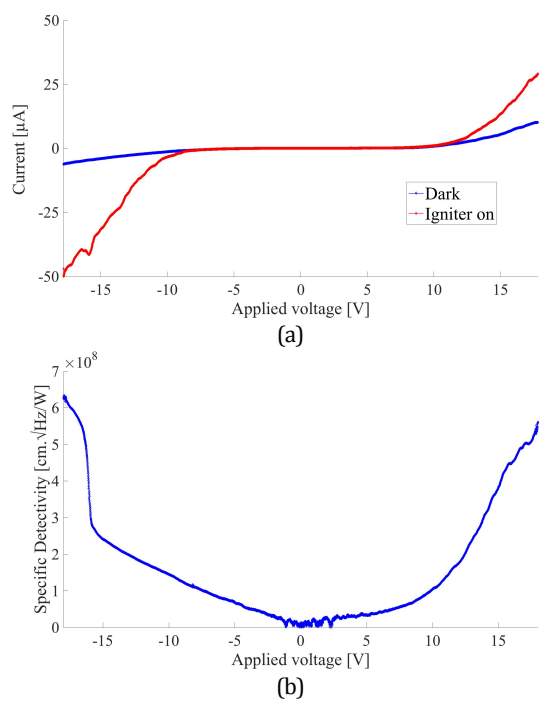

**Fig. 9.** (a) Measured current-voltage characteristic of array for infrared detection. (b) Specific detectivity as a function of applied voltage based on current-voltage characteristic curves.

Since the electron emission mechanism is facilitated by applying higher DC voltages photocurrent and specific detectivity also increase with voltage. The asymmetry in current-voltage characteristics and consequently in the specific detectivity curve in positive and negative bias is due to physical asymmetry in the fabricated arrays. Avoiding semiconductors as the medium of charge transport helps to reduce the noise due to generation and recombination processes and enhance sensitivity. It is possible to tailor the design for spectral tunability based on the geometrical dependence. This allows fabrication of dual and multi-color devices. The performance of the resonant surface is not strongly contingent on cryogenic cooling to low temperatures. Although lower temperatures may still provide superior sensitivity, the contribution of noise caused by thermal effects is not the dominant mechanism in the signal to noise ratio. Specific detectivity of available uncooled detectors including InSb, PbSe, and InAs sensors operating in photovoltaic, photoconductive, and photoelectromagnetic modes and bolometers are shown to be on the order of $10^8$-$10^9 \frac{cm\sqrt{Hz}}{W}$ [33,38,39]. Although there are higher detectivities reported for InGaAs and HgCdTe sensors, the complications in processing of these materials can make the proposed plasmonic nano-array approach an alternative solution for infrared photodetection.

In summary, this work demonstrates enhanced photocurrents in a gold array supporting surface plasmons at near-infrared and infrared frequency ranges. We took advantage of the nonlinear light-matter interaction at the interface to significantly reduce the electrical and optical powers required for carrier generation. We showed that the photoemitted current scales with the array area. The geometrical degree of freedom allows for spectral tunability of the optical response. Using a resonant metal structure as the active detection medium rather than semiconductors provides high specific detectivity that is on par with the best existing uncooled semiconductor infrared photodetectors.

**Funding Information.** Defense Advanced Research Projects Agency (DARPA) (N00014-13-1-0618)

**Acknowledgment**. The authors thank UC San Diego nanofabrication facility staff including S. Parks, L. Grissom, R. Anderson, I. Harris, and X. Lu for the helpful discussions, and especially Dr. M. Montero for performing E-beam lithography exposures.